\documentclass[prc,preprintnumbers,showpacs,twocolumn,amsmath,amssymb]{revtex4}
\usepackage{graphicx}
\usepackage{longtable}
\usepackage{dcolumn}
\begin{document}

\title{{\Large Small Quadrupole Deformation for the Dipole Bands in $^{112}$In}}

\author{\large T. Trivedi$^{1}$}
\author{\large R. Palit$^{1}$}
\email[Corresponding Author: ]{palit@tifr.res.in}
\author{\large J. Sethi$^{1}$}
\author{\large S. Saha$^{1}$}
\author{\large S. Kumar$^2$}
\author{\large Z. Naik$^{1}$}
\author{\large V. V. Parkar$^{1}$\footnote{Present address: Departamento de F\'{i}sica Aplicada, Universidad de Huelva, E-21071 Huelva, Spain}}
\author{\large B.S. Naidu$^{1}$}
\author{\large A.Y. Deo$^{1}$}
\author{\large A. Raghav$^{1}$}
\author{\large P. K. Joshi$^1$}
\author{\large H. C. Jain$^{1}$}
\author{\large S. Sihotra$^3$}
\author{\large D. Mehta$^3$}
\author{\large A. K. Jain$^{4}$}
\author{\large D. Choudhury$^{4}$}
\author{\large D. Negi$^{5}$}
\author{\large S. Roy$^{6}$}
\author{\large S. Chattopadhyay$^{6}$}
\author{\large A.K. Singh$^{7}$}
\author{\large P. Singh$^{7}$}
\author{\large D.C. Biswas$^{8}$}
\author{\large R.K. Bhowmik$^{5}$}
\author{\large S. Muralithar$^{5}$}
\author{\large R. P. Singh$^{5}$}
\author{\large R. Kumar$^{5}$}
\author{\large K. Rani$^{5}$}

\affiliation{$^1$Department of Nuclear and Atomic Physics,
Tata Institute of Fundamental Research, Mumbai - 400005, INDIA}
\affiliation{$^{2}$Department of Physics and Astrophysics,
University of Delhi, Delhi-110007, INDIA}
\affiliation{$^3$Department of Physics, Panjab University,
Chandigarh - 160014, INDIA}
\affiliation{$^4$Department of Physics, Indian Institute of Technology,
Roorkee - 247667, INDIA}
\affiliation{$^{5}$Inter University Accelerator Centre, New
Delhi-110067, INDIA}
\affiliation{$^6$Saha Institute of Nuclear Physics, I/AF, Bidhannagar,
Kolkata - 700064, INDIA}
\affiliation{$^7$Department of Physics $\&$ Meteorology, Indian Institute of Technology Kharagpur, Kharagpur - 721302, INDIA}
\affiliation{$^8$Nuclear Physics Division, Bhabha Atomic Research Centre, Mumbai - 400085, INDIA}

\begin{abstract}
High spin states in $^{112}$In were investigated using $^{100}$Mo($^{16}$O,
p3n) reaction at 80 MeV. The excited level have been observed up to 5.6 MeV excitation energy and spin $\sim$ 20$\hbar$ with the level scheme showing three dipole bands. The polarization and lifetime measurements were carried out for the dipole bands. Tilted axis cranking model
calculations were performed for different quasi-particle configurations of
this doubly odd nucleus. Comparison of the calculations of the model with the
B(M1) transition strengths of the positive and negative parity bands firmly
established their configurations.
\end{abstract}


\maketitle

\section {Introduction}
Various nuclear excitation modes have been understood by considering the symmetry of
nuclear mean field and relative orientation of the total angular momentum with
respect to its principal axes. In particular, the investigation of generation of high
angular momentum states in nuclei based on symmetry consideration and geometrical models has been extremely successful in case of novel excitation modes like magnetic, anti-magnetic and chiral rotations \cite{frau}. Studies of high spin
states in Ag, Cd and In near $A \sim 110$ isotopes continue to reveal new
aspects of these modes of excitations in nuclei \cite {si03,zh01,da05,ch10,
ro11}. Magnetic rotational (MR) bands were reported in
$^{108,110}$In isotopes \cite {ch01} and recently, two dipole
bands observed in $^{106}$In isotope were explained in terms of different K
values with same quasiparticle orbitals \cite {de09,pa10}.
The nuclei in this region exhibit many exciting features involving regular
band structures arising from the occupancy of the valance protons and neutrons
in g$_{9/2}$ and h$_{11/2}$ orbitals, respectively. Such high-$j$ orbitals
are now well known for generation of rotation like sequences of M1 transitions
called shears bands \cite{rmc}. Another interesting aspect of nuclei in this mass region is the
appearance of  $\Delta I = 1$ doublet bands \cite{104Rh,prm,ko10} with same parity,
which are nearly degenerate. In this picture, the degenerate bands observed in the lab-frame arise in
nuclei due to the possibility of forming mutually perpendicular coupling of three angular momenta of the
collective triaxial core, valence neutron and proton either in left or right
handed system in the intrinsic frame of the nucleus. Relativistic Mean Field (RMF)
calculations have been reported for the odd-odd nuclides in the $A\sim100$ mass
region \cite {me06}. Favorable $\gamma$ deformation required for chirality has
been predicted in $^{102-110}$Rh, $^{108-112}$Ag, and $^{112}$In odd-odd isotopes.
These triaxial doubly odd isotopes are predicted to have multiple chiral bands.
Recently, $^{112}$In has attracted considerable experimental attention and an elaborate level scheme of $^{112}$In has
been reported in Ref \cite {he10} based on $^7$Li + $^{110}$Pd reaction.
One of the motivation of the present work is to assign the parity of different dipole bands through
polarization measurements for detailed understanding of their configurations.
In addition, a heavier ion beam was chosen to populate excited levels of $^{112}$In with higher recoil
velocity required for lifetime measurements using Doppler-shift attenuation method (DSAM).
The results of polarization and lifetime measurements have been used along with tilted axis cranking (TAC)
calculations \cite {frau93,ne10} to obtain the shape parameters and quasiparticle configurations for different bands in $^{112}$In.
The present experimental details and results are discussed in section II and III.

\begin{figure*}[ht]
\begin{center}
\includegraphics[scale=0.5, angle = -90]{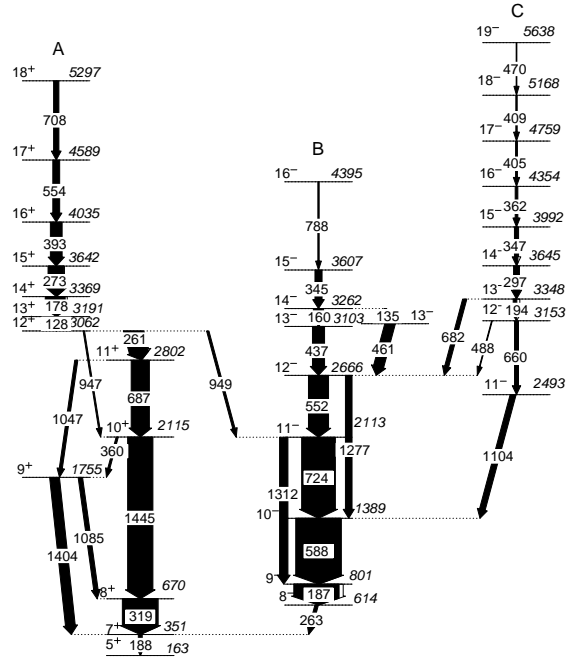}%
\caption{\label{fig1} Partial level scheme of $^{112}$In relevant for the
present work. The $\gamma$-rays energies are in keV.}
\end{center}
\end{figure*}

\begin{figure*}
\begin{center}
\includegraphics[scale=0.571, angle=-90]{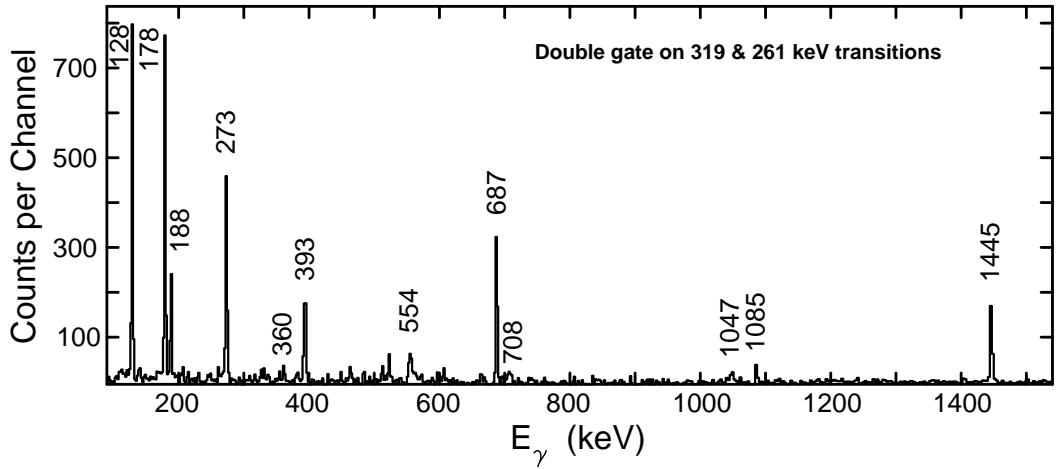}%
\caption{\label{fig2}Double gated spectrum with 319 and 261-keV transitions indicating all the dipole transitions of Band A.}
\end{center}
\end{figure*}

\begin{figure*}
\begin{center}
\includegraphics[scale=0.571, angle=-90]{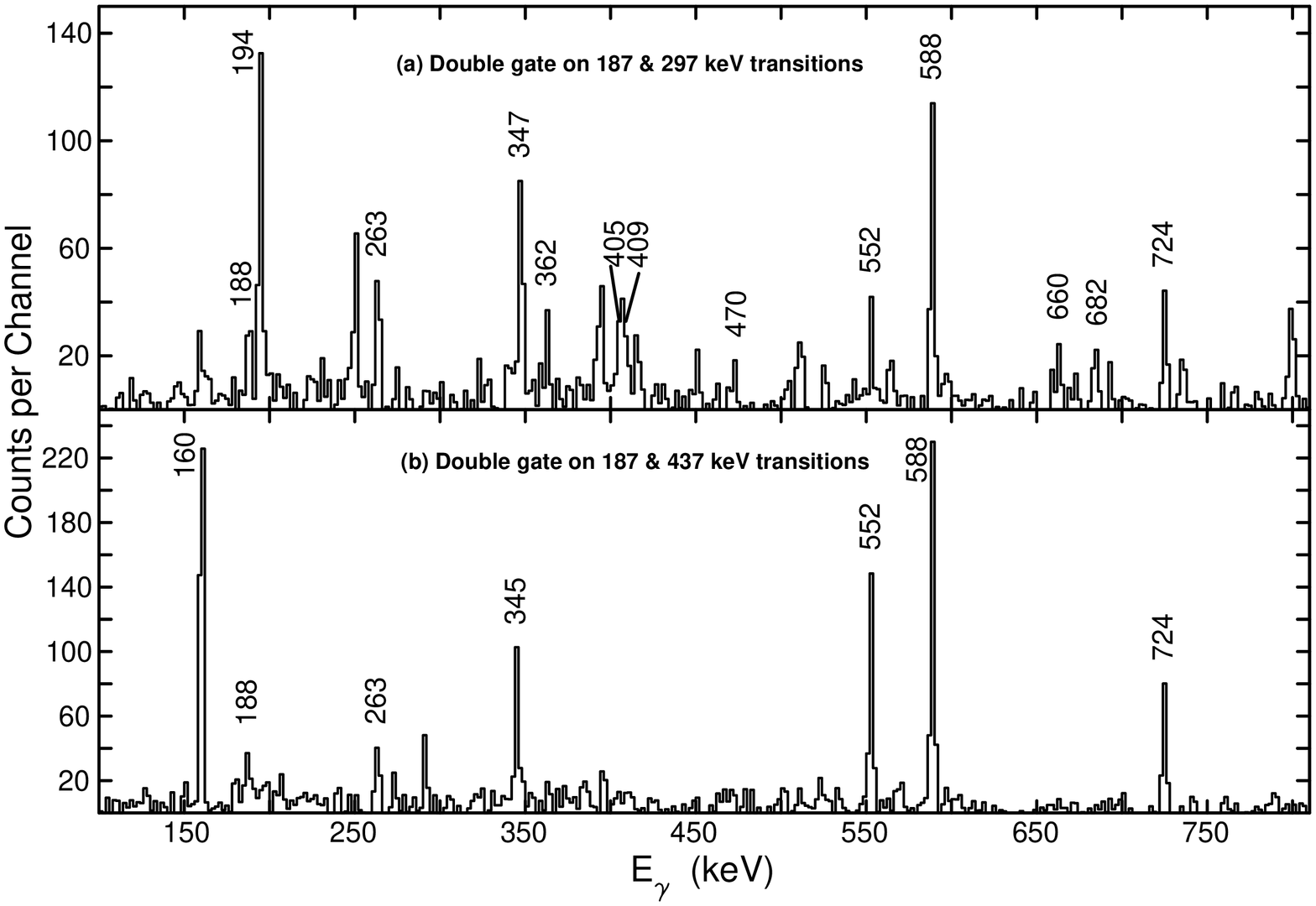}%
\caption{\label{fig3}Double gated spectra obtained by (a)
gates on 187 and 297-keV, and (b) 187 and 437-keV transitions. $\gamma$-rays transitions associated
with $^{112}$In are labeled with their energies in keV.}
\end{center}
\end{figure*}

\begin{figure*}
\begin{center}
\includegraphics[scale=0.571, angle=-90]{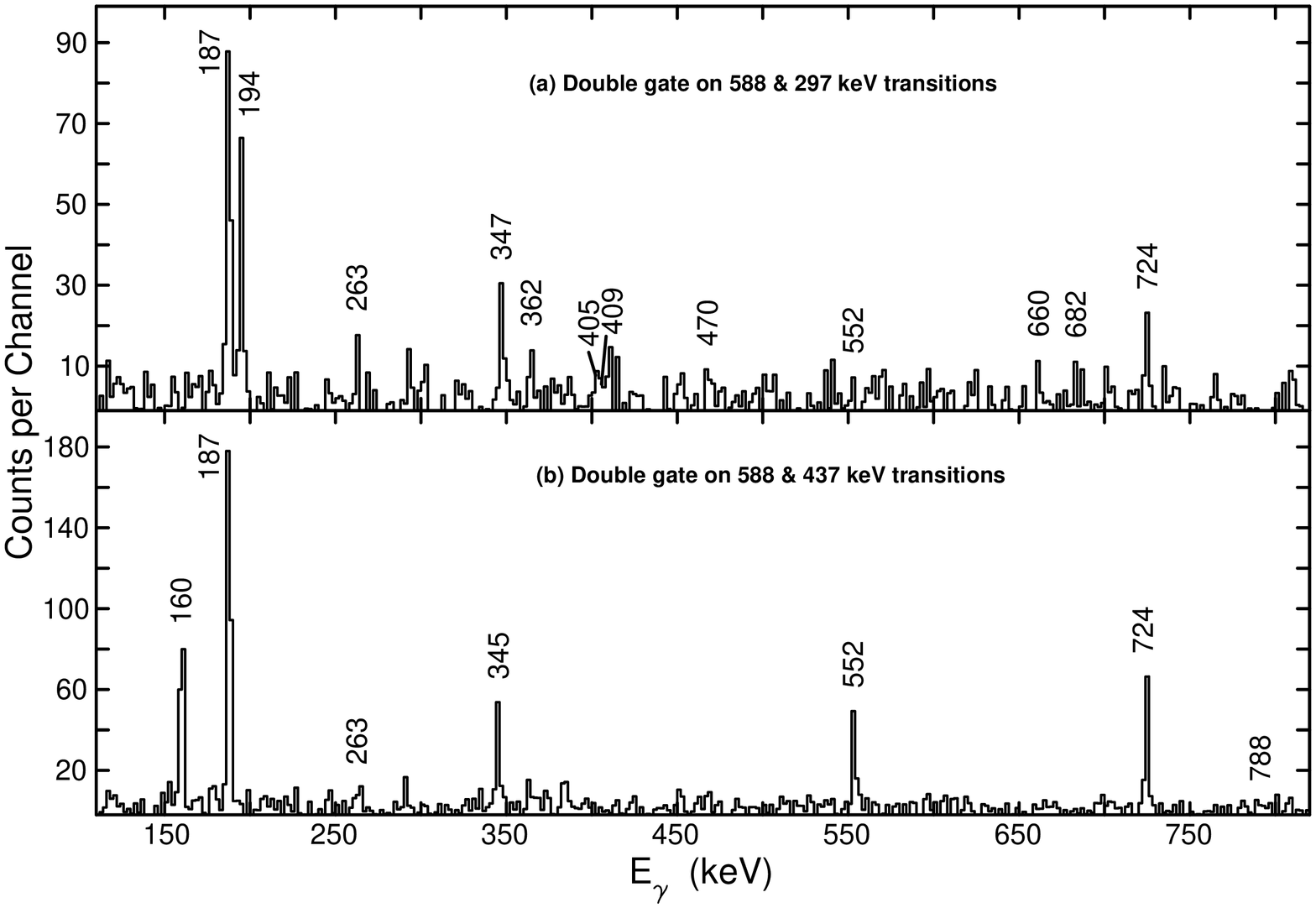}%
\caption{\label{fig4}Double gated spectra obtained by (a)
gates on 588 and 297-keV, and (b) 588 and 437-keV transitions. $\gamma$-rays transitions associated
with $^{112}$In are labeled with their energies in keV.}
\end{center}
\end{figure*}
\section {Experimental Details and Analysis Procedure}

High spin states in $^{112}$In were populated using $^{100}$Mo($^{16}$O, p3n)
reaction. The 80-MeV $^{16}$O beam was obtained from 15-UD Pelletron accelerator at
IUAC, New Delhi. The target consisted of 2.7 mg/cm$^{2}$ $^{100}$Mo with backing of $\sim$ 12
mg/cm$^2$ Pb to stop the recoiling ions produced in the  reaction.
Indian National Gamma Array (INGA) consisting of
eighteen Compton suppressed clover detectors was used to detect $\gamma$-rays emitted in the reaction.
This collaborative research facility was initiated by Tata Institute of Fundamental Research, IUAC, Bhabha Atomic Research Centre, 
Saha Institute of Nuclear Physics, Variable Energy Cyclotron Centre, UGC-DAE-Consortium for Scientific Research, and many 
Universities in India.
The clover detectors were arranged in five rings, at 32$^{\circ}$, 57$^{\circ}$,
90$^{\circ}$, 123$^{\circ}$, and 148$^{\circ}$ with respect to the beam direction \cite {mu10,bh01}.
The data were acquired when at least three clover detectors fired simultaneously.
$^{133}$Ba and $^{152}$Eu radioactive sources were used for the energy calibration and determination
of relative photopeak efficiency of the array.
After gain matching of individual crystals, add-back spectra were generated for
all the clovers and the coincidence data was stored in the $\gamma-\gamma$ matrix
which has about $ 1.1 \times 10^9$ events in total. An $E_\gamma\times E_\gamma\times E_\gamma$ cube was also constructed from the data. The RADWARE software package \cite {rad95} was used for the analysis of these matrices and the cube. The partial level scheme of $^{112}$In is shown in Fig. \ref{fig1}.
Various gated spectra relevant for identifying transitions in bands A, B and C are shown in Figs. [\ref{fig2}-\ref{fig4}].
The double gated spectrum obtained in coincidence with 319 and 261 keV transitions shown in Fig.\ref{fig2} depicts the 128 - 178 - 273 - 393 - 554 - 708-keV cascade of band A. Similarly, the four double gated spectra given in Fig.\ref{fig3} and \ref{fig4}
with gate on 187-297 keV, 187 - 437 keV, 588 - 297 keV and 588 - 437 keV
transition-pairs show the $\gamma$-rays of bands B and C as given in Refs.\cite {he10,he09}.

\begin{figure}
\begin{center}
\includegraphics[scale=0.481]{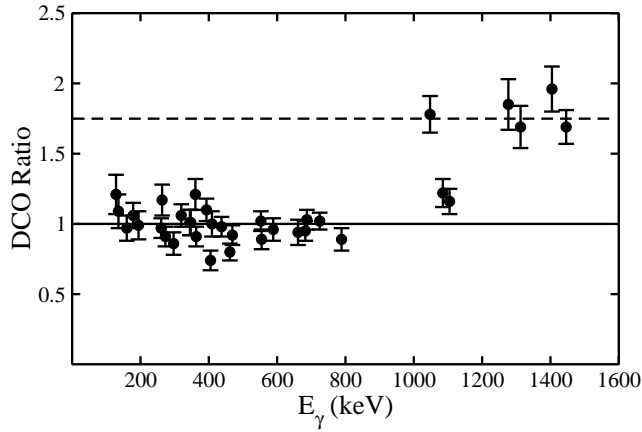}%
\caption{\label{fig5}Plot for DCO ratios of different transitions.}
\end{center}
\end{figure}

\begin{figure}
\begin{center}
\includegraphics[scale=0.481]{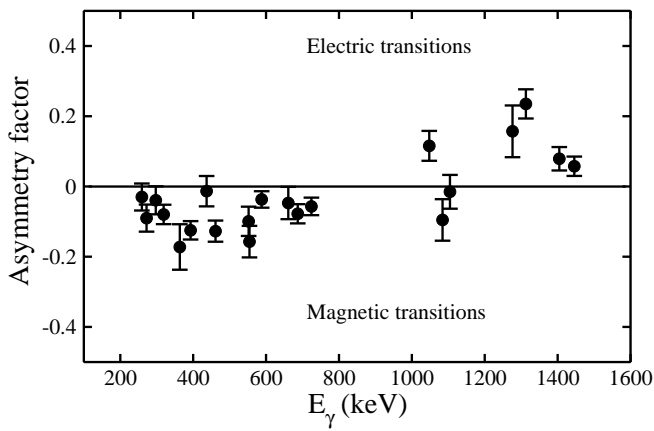}%
\caption{\label{fig6}Plot for polarization asymmetry for different transitions.}
\end{center}
\end{figure}

\begin{figure*}
\begin{center}
\includegraphics[scale=0.481,angle=0]{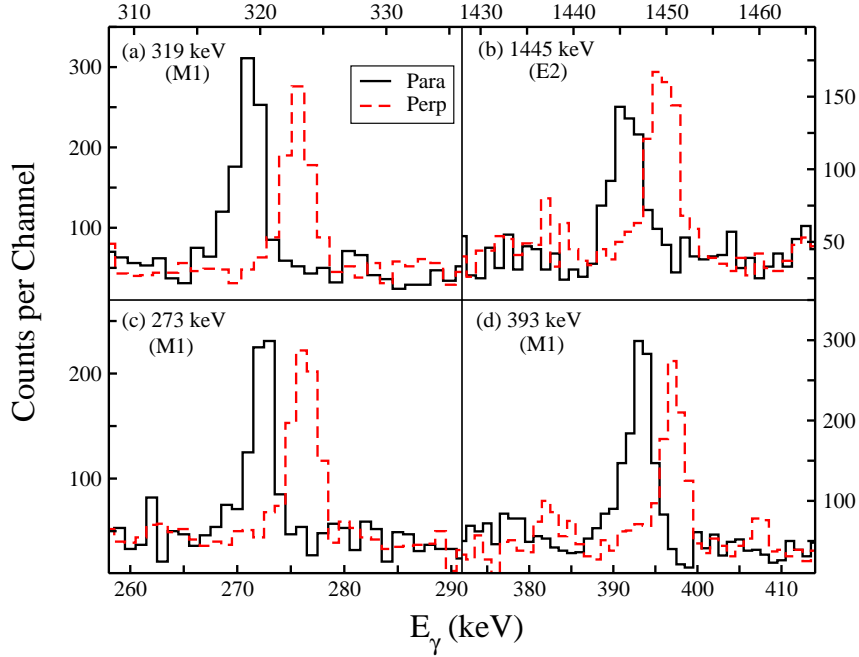}%
\caption{\label{fig7}(Color online) 261-keV gated spectra of the perpendicular
and parallel Compton scattering in the $90^\circ$ clover detectors 
corresponding to (a) 319-keV, (b) 1445-keV, (c) 273-keV, and (d) 393-keV 
transitions. Higher counts for 273, 319 and 393-keV transitions in parallel 
scattered spectrum indicates their magnetic character, while (d) suggests 
electric nature for 1445-keV transition. An offset of 4 keV has been introduced
between the parallel and perpendicular spectra for clarity.}
\end{center}
\end{figure*}

\begin{table*}
\caption{Excitation energies, $\gamma$-ray energies, intensities, DCO-ratio, multipolarity, IPDCO value and initial and final state spin of the transitions of $^{112}$In deduced from the present work are listed. The uncertainties in the energies of $\gamma$-rays are 0.3 keV for intense peaks and 0.7 keV for weak peaks.}
\begin{ruledtabular}
\begin{tabular}{ccccccc}

\text{$E_{x}$ (keV)} & \text{$E_{\gamma}$ (keV)}  & \text{$I_{\gamma}$} & \text {$R_{DCO}(D)$} & \text{$Multipolarity$} & \text{$IPDCO$} & \text{$I_i^{\pi} \rightarrow I_f^{\pi}$}  \\
\hline
\hline
614 &263.1& &1.17(10) & & &${8}^{-}\rightarrow{7}^{+}$ \\
670 &319.2&104.3(8) &1.06(7) &$\emph{M}1$ &-0.079(28) &${8}^{+}\rightarrow{7}^{+}$ \\
801 &187.1& & & & & ${9}^{-}\rightarrow{8}^{-}$\\
1389 &588.2&134.8(9) &0.96(7) &$\emph{M}1$ &-0.037(23) & ${10}^{-}\rightarrow{9}^{-}$\\
1755 &1084.8&13.1(1) &1.22(10) &$\emph{M}1$ &-0.095(59) & ${9}^{+}\rightarrow{8}^{+}$\\
     &1404.0&29.1(13) & 1.96(15)&$\emph{E}2$ &0.079(33) & ${9}^{+}\rightarrow{7}^{+}$ \\
2113 &724.3&100(6) &1.02(6)  &$\emph{M}1$ &-0.057(25) &  ${11}^{-}\rightarrow{10}^{-}$ \\
     &1312.5&24.8(4) &1.69(14)&$\emph{E}2$ &0.235(42) &  ${11}^{-}\rightarrow{9}^{-}$ \\
2115 &1445.2&75.1(4) &1.69(12) &$\emph{E}2$ &0.058(27) &  ${10}^{+}\rightarrow{8}^{+}$\\
     &360.4 &6.1(1) &1.21(11) & & & ${10}^{+}\rightarrow{9}^{+}$ \\
2493 &1104.2&17.2(2) &1.16(9) &$\emph{M}1$ &-0.015(48) & ${11}^{-}\rightarrow{10}^{-}$ \\
2666 &552.4&59.4(4) &1.02(6) &$\emph{M}1$ &-0.099(42) &  ${12}^{-}\rightarrow{11}^{-}$\\
     &1276.7&19.6(2) &1.85(18) &$\emph{E}2$ &0.157(73) &  ${12}^{-}\rightarrow{10}^{-}$\\
2802 &686.9&53.9(3) &1.03(7) &$\emph{M}1$ &-0.078(27) & ${11}^{+}\rightarrow{10}^{+}$ \\
     &1047.4&9.1(1) &1.78(13) &$\emph{E}2$ &0.116(42) & ${11}^{+}\rightarrow{9}^{+}$ \\
3062 &260.6&56.9(2) &0.97(7) &$\emph{M}1$ &-0.030(38) & ${12}^{+}\rightarrow{11}^{+}$ \\
     &947.4&4.2(2) & & & &  ${12}^{+}\rightarrow{10}^{+}$\\
     &949.1&6.8(3) & & & &  ${12}^{+}\rightarrow{11}^{-}$\\
3103 &437.1&35.6(3) &0.98(6) &$\emph{M}1$ &-0.013(43) & ${13}^{-}\rightarrow{12}^{-}$ \\
3127 &461.4&30.3(2) &0.80(5) &$\emph{M}1$ &-0.127(30) & ${13}^{-}\rightarrow{12}^{-}$\\
3153 &660.2&11.7(3) &0.94(9) &$\emph{M}1$ &-0.047(46)  & ${12}^{-}\rightarrow{11}^{-}$\\
     &487.7&2.2(1) & & & & ${12}^{-}\rightarrow{12}^{-}$ \\
3191 &128.3&60.1(2)&1.21(13)& & & ${13}^{+}\rightarrow{12}^{+}$ \\
3262 &159.6&26.7(2) &0.97(9) & & & ${14}^{-}\rightarrow{13}^{-}$ \\
     &135.3&6.1(1) &1.09(11) & & & ${14}^{-}\rightarrow{13}^{-}$ \\
3348 &194.2&10.8(1) &0.99(9) & & & ${13}^{-}\rightarrow{12}^{-}$ \\
     &681.9&12.1(2) &0.95(6) & & &  ${13}^{-}\rightarrow{12}^{-}$\\
3369 &178.5&59.3(2)&1.06(9) &  & & ${14}^{+}\rightarrow{13}^{+}$\\
3607 &344.6&20.7(1) &1.01(9) & & & ${15}^{-}\rightarrow{14}^{-}$ \\
3642 &272.7&47.2(2) &0.91(6) &$\emph{M}1$ &-0.090(39) & ${15}^{+}\rightarrow{14}^{+}$ \\
3645 &296.9&17.0(2) &0.86(7) &$\emph{M}1$ &-0.039(40) & ${14}^{-}\rightarrow{13}^{-}$ \\
3992 &347.1&10.4(1)&1.01(9) & & & ${15}^{-}\rightarrow{14}^{-}$ \\
4035 &393.3&34.5(2) &1.10(7) &$\emph{M}1$ &-0.139(26) &  ${16}^{+}\rightarrow{15}^{+}$\\
4354 &362.4&7.8(1) &0.91(7) &  & &   ${16}^{-}\rightarrow{15}^{-}$\\
4395 &787.9&3.8(1) &0.89(7) &  & &   ${16}^{-}\rightarrow{15}^{-}$\\
4589 &554.2&20.3(2) &0.89(6) & $\emph{M}1$ &-0.157(44) &  ${16}^{+}\rightarrow{15}^{+}$\\
4759 &404.7&5.4(1) &0.74(7) & & & ${17}^{-}\rightarrow{16}^{-}$ \\
5168 &409.2&3.9(1) &1.00(9) & & & ${18}^{-}\rightarrow{17}^{-}$ \\
5297 &707.6&15.6(1) & & & &  ${18}^{+}\rightarrow{17}^{+}$\\
5638 &470.0&2.0(1) &0.92(7) & & & ${19}^{-}\rightarrow{18}^{-}$ \\

\end{tabular}
\end{ruledtabular}
\label{tab1}
\end{table*}

The directional correlation of oriented states (DCO) and integrated
polarization direction correlation (IPDCO) analysis were carried out to
determine the spin and parity of different states.
The multipolarity of $\gamma$-rays were deduced from the angular correlation
analysis \cite {Kr89} using the method of directional correlation from oriented states
ratios (DCO) of two coincident gamma rays $\gamma_1$ and $\gamma_2$ , given
by:

\begin{eqnarray*}
R_{DCO}  = \frac
{I\gamma_1\mbox{ observed at 32}^\circ\mbox{ gated on }\gamma_2\mbox{ at 90}^\circ}
{I\gamma_1\mbox{ observed at 90}^\circ\mbox{ gated on }\gamma_2\mbox{ at 32}^\circ}
\end{eqnarray*}

In the present geometry of detectors, the DCO ratios obtained with a stretched quadrupole (dipole) gate are 0.5(1.0) and 1.0(2.0) for the pure dipole and quadrupole transitions, respectively. The DCO ratios obtained are shown in Fig. \ref{fig5}. The
extracted DCO values were obtained with gate on strong $\Delta I = 1$
dipole transitions.

The clover detectors at 90$^{\circ}$ were used as a Compton-polarimeter which
helps in identifying the electric or magnetic nature of $\gamma$-rays \cite {St99, Pa00}.
For a Compton-polarimeter, polarization asymmetry $\Delta$ of the transition is defined as \cite {St99}
\begin{equation}\label{ratio}
 \Delta=\frac{a(E_\gamma)N_\perp-N_\parallel}{a(E_\gamma)N_\perp+N_\parallel}
\end{equation}
where, $N_\perp$($N_\parallel$) is the number of counts of $\gamma$ transitions scattered perpendicular (parallel) to the reaction plane.
The correction factor $a(E_\gamma)$ is a measure of the perpendicular to parallel scattering asymmetry within the crystals of clover.
For the $90^\circ$ degree detectors this parameter has been found to be 0.98(1) from the analysis of decay data of the $^{152}$Eu radioactive source.
For linear polarization measurement, two asymmetric matrices corresponding to parallel and
perpendicular segments of clover detectors (with respect to the emission plane)
along one axis and the coincident $\gamma$-rays along the another axis were
constructed \cite{la05}. Then integrated polarization direction correlation
(IPDCO) analysis was carried out. A positive value of IPDCO ratio indicates an
electric transition while negative value for a magnetic transition.
The positive and negative asymmetry parameters of different transitions depicted in Fig. \ref{fig6} indicate their electric
and magnetic nature, respectively.
The 261 keV gated spectra generated from parallel ($N_\parallel$ ) and perpendicular ($a(E_\gamma ) N_\perp$) scattering events
observed in the 90$^\circ$ clover detectors are shown in Fig. \ref{fig7}.
The higher counts of the 273, 319 and 393-keV transitions in the parallel 
scattering spectrum compared to that in perpendicular scattering spectra 
suggested their magnetic nature, while the reverse nature of the spectra for 
1445 keV transition confirms its electric multipolarity.

\begin{figure*}
\begin{center}
\includegraphics[scale=.48]{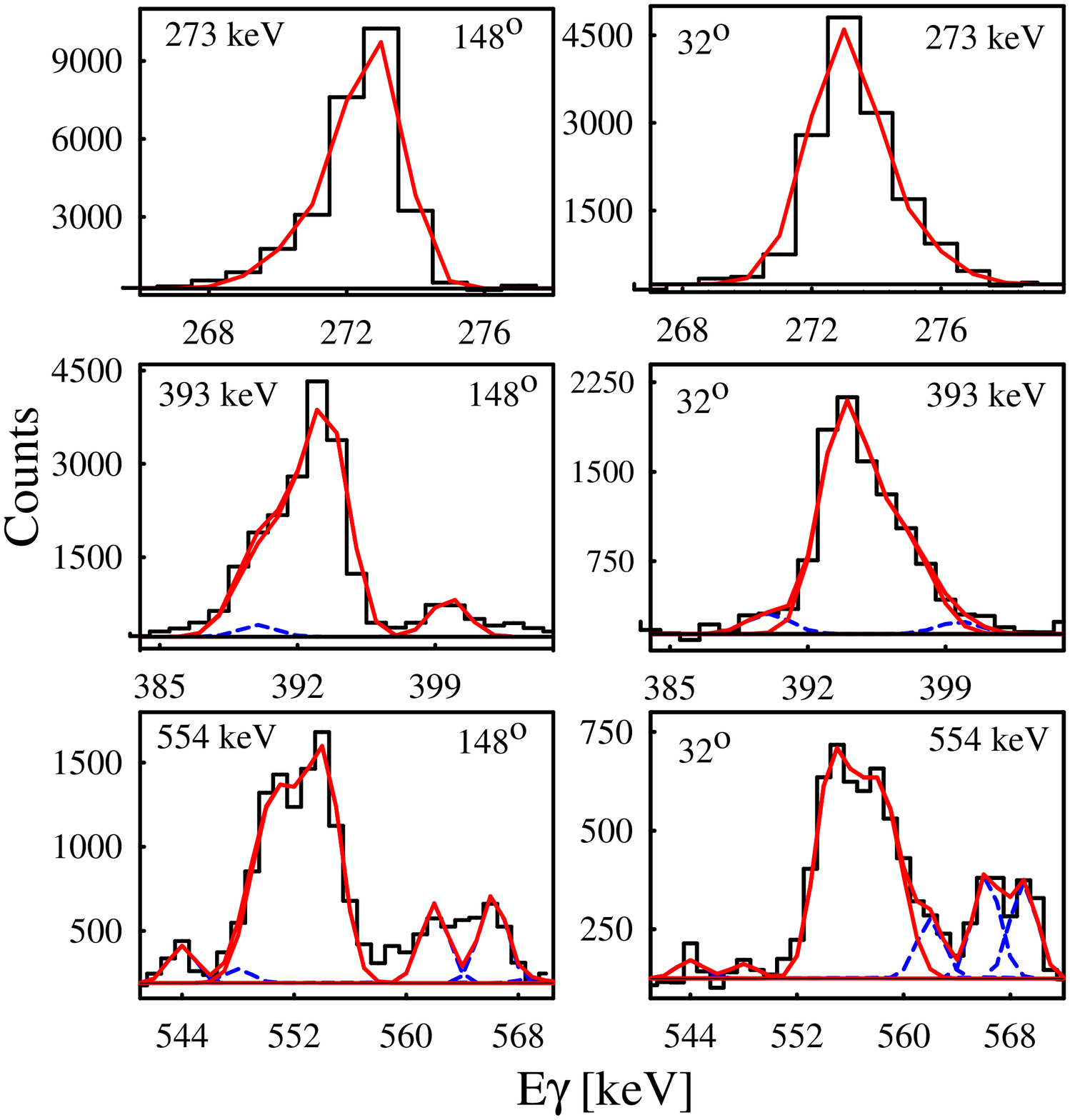}%
\caption{\label{fig8}
(Color online) Representative spectra (with gate on the
lower transition) along with theoretically fitted line shapes for
273, 393 and 554-keV transitions in the positive-parity yrast
band of $^{112}$In for $\gamma$-ray spectra at 32$^\circ$, and 148$^\circ$ with
respect to the beam direction. The contaminant peaks are marked by dashed lines.}
\end{center}
\end{figure*}

For the application of Doppler-shift attenuation method, line shapes were obtained from the
background-subtracted spectra projected from the two matrices consisting of
events in the 148$^\circ$ or 32$^\circ$ detectors along one axis and
all other detectors along the second axis, respectively. These matrices
contained approximately 5.8 $\times$10$^8$ and 4.0$\times$10$^8$ coincidence
events, respectively. The forward and backward Doppler shifted line shapes of
273, 393 and 554-keV transitions were shown in Fig. \ref{fig8}.
The line shape spectra were generated by putting gate on transition below the level of interest.

\section{Analysis Results}

\subsection{DCO and Polarization measurements}
The DCO values of the transitions from the higher levels are obtained with gate on 319-keV transition.
The polarization asymmetry values of the 319, 1445 and
687-keV transitions confirm the positive parity for the $11^+$ level at
2802 keV excitation energy. The 128 - 178 - 273 - 393 - 554 - 708-keV cascade
present in the 319 and 261 keV double gated spectrum shown in Fig. \ref{fig2} feeds
the 2802 keV level through 261-keV transition. Polarization and DCO values
obtained for 261 keV transition establishes the positive parity of the band A. The polarization asymmetry
for 273, 393 and 554-keV transitions are found to be negative suggesting their
magnetic character. The IPDCO of low energy transitions with energy 128 and 178-keV transitions could not be extracted and assumed to be magnetic based on the systematics of dipole bands in this mass region.
Further, E2/M1 multipole mixing ratio analysis \cite {Kr89} was carried out
for the dipole transitions of band A, viz, 273, 393, and
554, from their extracted $R_{DCO}$ values with gate on 319-keV pure
$\Delta I = 1$ transition.
The measured $R_{DCO}$ ratios suggest that these $\gamma-$rays
have a small E2/M1 multipole mixing ratio $ -7^0 < \arctan(\delta) < -1^0$. This
along with the IPDCO measurements suggest pure M1 nature for these
intraband $\Delta I = 1$ transitions.

The transitions of bands B and C were obtained from the double gated spectra given in Figs. \ref{fig3} and \ref{fig4}.
Based on the double gated spectra obtained with 187 - 437 keV and 588 - 437 keV pair of transitions suggest placement of 160 - 345 - 788-keV cascade above the 3103 keV state with $I^\pi$ = $13^-$. The transitions placed in band C are shown in Figs. \ref{fig3} and \ref{fig4} in the spectra obtained with 187 - 297 and 588 - 297 keV pairs. The band-head of band C was assigned a negative parity with spin $11^-$ due to the measured DCO and IPDCO of the 1104 and 660 keV transitions. Two more inter-band transitions between bands B and C with energy 488 and 682 keV are also observed in the gated spectrum. The 194 - 297 - 347 - 362 - 405 - 409 - 470-keV cascade was reported in \cite {he09}. This band C has a lower yield in the present reaction compared to band A. Therefore, only the IPDCO of 297-keV transition could be extracted and found to have magnetic nature. The DCO ratio for the 194, 297, 347 and 362-keV transitions are found to be around 1.0 suggesting the $\Delta I = 1$ for these transitions.


\begin{table*}
\caption{\label{tab:table2}Lifetimes of the states and reduced B($M1$) strengths of transitions of $^{112}$In are listed.}
\begin{ruledtabular}
\begin{tabular}{ccccc}
$I^{\pi}_{i}$ & $E_{\gamma, M1}$ (keV) & $\tau$ (ps) & $B(M1)$ $(\mu_{N})^{2}$ \\
\hline
${15}^+$ & 272.7  & $0.83^{+0.04}_{-0.04}$ & $3.28^{+0.14}_{-0.14}$   \\
${16}^+$ & 393.3  & $0.49^{+0.03}_{-0.03}$ & $1.89^{+0.11}_{-0.10}$ 	\\
${17}^+$ & 554.2  & $0.22^{+0.02}_{-0.03}$ & $1.52^{+0.20}_{-0.14}$   \\
${18}^+$ & 707.6  & $<0.25$ & $>0.63$   \\
${15}^-$ & 347.1  & $0.72^{+0.25}_{-0.16}$ & $1.87^{+0.55}_{-0.48}$  \\
${16}^-$ & 362.4  & $<0.60$ & $>1.97$  \\
\end{tabular}
\end{ruledtabular}
\end{table*}


\subsection{Lifetime measurements}

Lifetimes of the states of band A and C have been measured by DSAM.
In the analysis, gating transitions were below the transitions of interest.
For analyzing the line shapes of different transition of $^{112}$In, LINESHAPE \cite{wel91} program was used.
The program takes into account the energy loss of the beam through the target and the energy
loss and angular straggling of the recoils through the target
and the backing. For the energy loss calculations, we have
used the shell-corrected Northcliffe and Schilling stopping
powers \cite{nor70}. The value of the time step and the number
of recoil histories were chosen to be 0.01 ps and 5000,
respectively.
In the fitting procedure, program obtains a ${\chi}^2$ minimization of the
fit for transition quadrupole moments ($Q_t$) for the transition of interest,
transition quadrupole moments $Q_t$(SF) of the modeled side feeding cascade,
the intensity of contaminant peaks in the region of interest and the
normalizing factor to normalize the intensity of fitted transition.
The best fit was obtained through the least square minimization
procedures SEEK, SIMPLEX, and MIGRAD referred in \cite{wel91}

In band A, the Doppler-broadened line shapes were observed
for the 273, 393, 554 and 708-keV transitions above
I$^\pi$=14$^+$ state. The line shapes of these transition were
obtained by putting gate on 178-keV transition. Assuming 100$\%$ side feeding into the top of band,
an effective lifetime of the top-most state was estimated which
was then used as an input parameter to extract the lifetimes of
lower states in the cascade. The side feeding into each level of the band was
considered as a cascade of five transitions having a fixed
moment of inertia comparable to that of the in-band sequences.
The energies of $\gamma$-rays and side-feeding intensities
were used as input parameters for the line shape analysis. Side-feeding intensities
were calculated by using an asymmetric $\gamma$-$\gamma$ matrix comprising $\gamma$-rays detected
by detectors at 90$^\circ$ along one axis and all other detectors along the second axis.
Once the $\chi^2$ minimization was obtained by MINUIT\cite{Jam75} program,
the background and the contaminant peak parameters were fixed
and the procedure was followed for the next lower level.
After obtaining $\chi^2$ minimization for each level, a global fit
was carried out, with the background and the contaminant
peak parameters of all the levels kept fixed.
The side feeding lifetimes were found to be faster than the level lifetimes similar to the measurements reported in nearby nuclei in this mass region \cite {ch01,ne10}.

The final values of lifetimes were obtained by taking weighted averages of the results
obtained from the two separate fits which were performed at 32$^\circ$ and 148$^\circ$.
For each band, B(\emph{M}1) values were calculated from measured lifetimes using the following relationship:

\begin{eqnarray}
B(M1) = \frac{{0.05697}}{{E_\gamma ^3 (M1)\tau \left[ {1 + \alpha _t (M1)} \right]}}\left[ {\mu _N^2 } \right]
\end{eqnarray}

Where, E$_\gamma$ is transition energy in MeV, $\tau$ is partial
lifetime of the transition deduced from fitted line shape of the state
and $\alpha_t$ is total internal conversion coefficient of the transition, respectively.
These results are listed in Table II and fitting of the theoretical
line shapes with the experimental data for 273, 393 and
554-keV transitions are shown in Fig. \ref{fig8}. The errors quoted
in lifetimes does not include the systematic errors from
the uncertainty in stopping power, which can be as large
as 15$\%$. Similarly, line shapes for 347 and 362-keV transitions of bands C were observed. After the fitting of
calculated line shapes, the lifetime of the respective states are given in Table II.

\begin{figure}
\begin{center}
\includegraphics[scale=0.481]{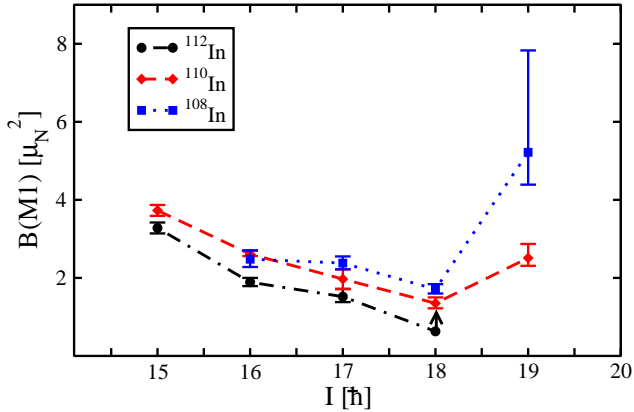}%
\caption{\label{fig9}
(Color online)Comparison of experimentally measured B(M1) transition strength as a function of spin
for dipole band 3 of $^{108}$In, $^{110}$In \cite{ch01} and band A of $^{112}$In.}
\end{center}
\end{figure}
\begin{figure*}
\begin{center}
\includegraphics[scale=0.551,angle=0]{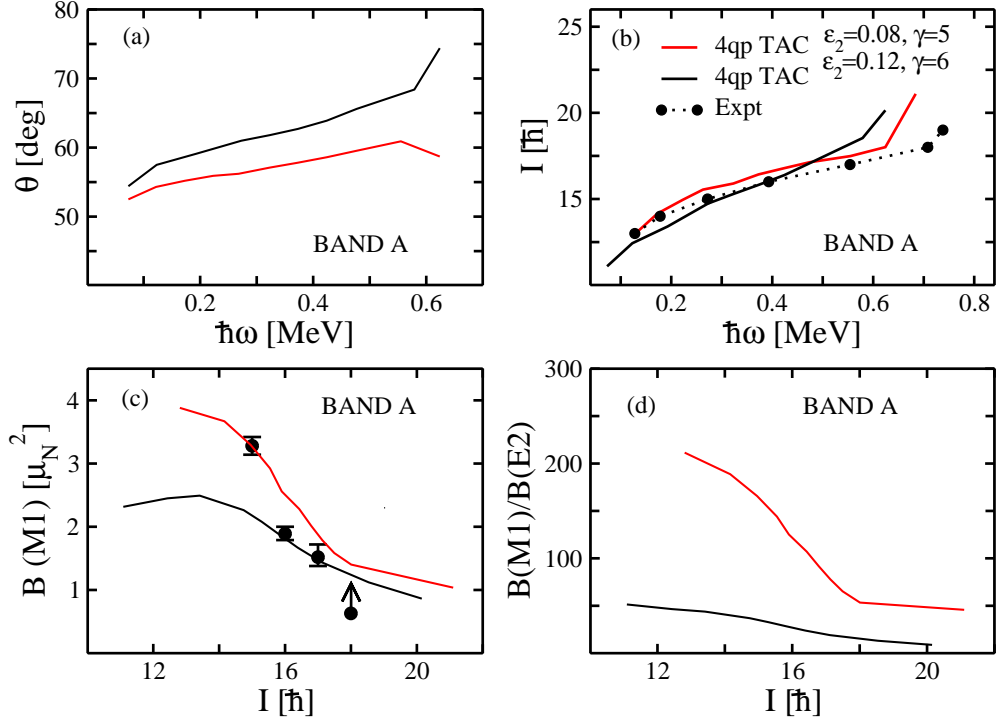}%
\caption{\label{fig10}(Color online) The results of TAC calculations for the positive parity band showing (a) the variation of tilt angle with rotational frequency, (b) spin (I)  {\it {vs.}} rotational frequency, (c) B(M1) transition strength {\it {vs.}} spin and (d) B(M1)/B(E2) ratio {\it {vs.}} spin. The experimental data shown as filled circle for band A of $^{112}$In are plotted in (b) and (c) for comparison.}
\end{center}
\end{figure*}

\begin{figure}
\begin{center}
\includegraphics[scale=0.551,angle=0]{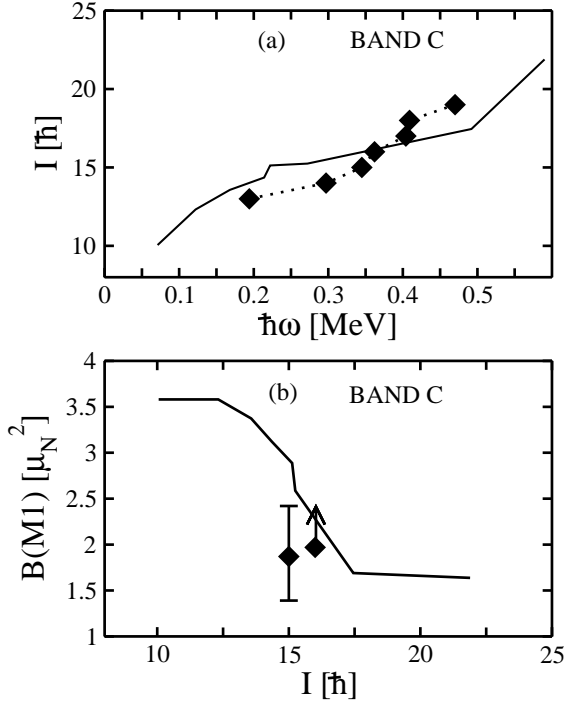}%
\caption{\label{fig11} The results of TAC calculations and its comparison with
the experimental data for negative parity band C of $^{112}$In.}
\end{center}
\end{figure}


\section{Comparison with model calculations}

The positive parity dipole band with band-head excitation energy 3.062 MeV has been observed up to $I^\pi=18^+$. Similar dipole bands have been observed in $^{108}$In and $^{110}$In with $\pi g_{9/2}^{-1} \otimes \nu (d_{5/2}/g_{7/2})(h_{11/2})^2$ quasi-particle configuration \cite{ch01}.
Fig. \ref{fig9} demonstrates the comparison of experimentally determined B(M1) transition strengths as a function of spin for dipole band of $^{112}$In with band 3 of $^{108}$In and $^{110}$In \cite{ch01} having almost similar configurations. It is evident that the B(M1) strength decreases with increasing spin up to
$I^\pi = 18^+$ in an identical way for all three dipole bands, which confirms 
the similar configuration for the positive parity dipole band of $^{112}$In. 
In addition, it is to be noted that the band crossing near $\hbar\omega = 0.6 $
MeV observed in these positive parity dipole bands of $^{108,110}$In is not 
seen in $^{112}$In \cite {ch01,he10}. Very recently, a positive 
parity dipole band with band head spin of $I^\pi=11^+$ has
been assigned the same configuration $\pi g_{9/2}^{-1} \otimes \nu (d_{5/2}/g_{7/2})(h_{11/2})^2$ in $^{114}$In \cite {li11}. This band has strong $\Delta I = 1$ transitions
with unobserved cross-over E2 transitions and no signature splitting similar
to the lighter odd-odd In isotopes. It will be interesting to perform lifetime 
measurements for the levels in this band of $^{114}$In to confirm the role of 
shear mechanism. 

In the discussion that follows, the experimental data of the dipole bands (labeled as A and C) are
compared with the results of TAC model calculations. The values of proton pairing gap parameter $\Delta_\pi = 0.99 $ MeV
and neutron pairing gap parameter $\Delta_\nu = 0.85$ MeV were used
in the TAC calculations. These values are 0.6 and 0.8 times
the odd and even mass difference, respectively. The chemical
potentials $\lambda_\nu$ (both proton and neutron) were chosen so that a particle number is conserved for Z = 49 and N = 63 . The values of deformation parameter $\epsilon_2 $ and $\gamma$
were obtained by Nilsson Strutinsky's minimization procedure
 \cite{25}. The nature of band A, which is a $\Delta I = 1$ positive-parity M1 band, was investigated.
A quasi-particle configuration $\pi g_{9/2}^{-1} \otimes \nu (d_{5/2}/g_{7/2})(h_{11/2})^2$ is used in tilted axis cranking calculations for the dipole band.
Similar, configurations have been assigned in lighter odd-odd In isotopes and supports magnetic rotations.
A minimum is found at deformation of $\epsilon_2 = 0.12$ and $\gamma = 6^\circ$. The calculated $ I~ vs.~ \hbar\omega$ and $B(M1) ~vs.~I$ plots
based on this global minima given in Fig. \ref{fig10} qualitatively explains the
measured data. 
However, the overall decreasing trend of the experimental B(M1) values in the 
measured range of spin is better explained by the results of the TAC calculations based on a staic deformation with $\epsilon_2 = 0.08$ and
$\gamma = 5^\circ$ .
In particular, the measured B(M1) value at lower spin ($I^\pi = 15^+$) 
is closer to the TAC calculation based on this lower deformation.
The static minima suggests almost a
constant tilt angle around $55^\circ$ and a large B(M1)/B(E2) values which
is in line with the non-observation of crossover E2 transitions in band A.
The low deformation of these states indicates that the contribution from
the core in angular momentum generation is negligible and the whole of
the angular momentum generation along the band can be attributed to the shears
mechanism. Similar situation have been reported for $^{107}$In \cite{ne10}and
$^{106,108}$Sn \cite{jen99}isotopes in the literature.
Moreover, the good agreement between the TAC calculations with constant tilt angle and
the measured variation of excitation energy and B(M1) values with spin for
the positive parity band firmly establishes magnetic rotation for positive
parity band A. This positive parity band
under consideration is probably one of the
ideal MR band due to very low deformation of 0.08. However, the confirmation
of the small deformation will require further
investigation of the crossover E2 transitions and measurement of the B(E2)
values. The band crossing observed at rotational frequency 0.6 MeV for 
$^{108,112}$In is due to lowering of the energy of 
$\pi g_{9/2}^{-1} \otimes \nu ((h_{11/2})^2(g_{7/2}/d_{5/2})^3$ configuration
at higher rotational frequency. Due to increasing neutron number up to N = 63
for $^{112}$In, $g_{7/2}$ orbitals are not available near the neutron Fermi surface for this configuration to be energetically favourable which explains the absence
of band crossing for the band A.

A negative parity dipole band C studied in the present work has excitation energy of 3153 keV.
The 194 - 297- 347 - 362 - 404 - 409 - 470-keV cascade extended the band C up to $I^\pi$ =  $19^-$. The
$\pi g_{9/2}^{-1} \otimes \nu (h_{11/2})^3$ configuration has been used in TAC for
comparison of the band C which gives a minimum with $\epsilon_2 = 0.13$ and $\gamma = 0^\circ$. The calculated $I~vs.~\omega$ plot (Fig. \ref{fig11}) explains the measured values resalable well. The measured B(M1) values at $I^\pi = 15^-, 16^-$ are also
reproduced with the results of TAC calculations.
This confirms the quasi-particle configuration of band C. In future, it will be interesting to perform the
recoil distance measurements (RDM) to determine the lifetime of lower
states of band C for testing the prediction of higher B(M1) values at lower spin states.

\section {Summary}

In summary, the polarization and lifetime measurements for the excited states
of the previously established level scheme of $^{112}$In have been carried out.
Out of the three dipole bands observed in the present experiment, band A is
found to have positive parity, while bands B and C have negative parity.
The extracted B(M1) values from the measured lifetime of the excited states of
band A has a decreasing trend with increasing spin. The TAC calculations based
on $\pi g_{9/2}^{-1} \otimes \nu (d_{5/2}/g_{7/2}) (h_{11/2})^2$ configuration reproduces
the measured trend of B(M1) with increasing spin. This establishes the shear
rotation in $^{112}$In for the positive parity dipole band. The
TAC calculations based on $\pi g_{9/2}^{-1} \otimes \nu (h_{11/2})^3$ quasi-particle configurations have been compared with the measurements for band C. The fair agreement of TAC calculations with the measurement suggests weak prolate deformation for the positive (A) and negative parity (C) dipole bands for $^{112}$In
contrary to the triaxial deformation as predicted in the RMF calculation 
\cite {me06}.
Further measurement of the crossover E2 transitions in band A would establish
this band as an ideal example of MR band owing to its small deformation of 0.08.

\section {Acknowledgments}
Authors would like to thank the IUAC pelletron staff for providing good
quality beam. The help and cooperation of the members of the INGA collaboration
for setting up the array is acknowledged. This work was partially funded by the Department of Science and Technology, Government of India (No. IR/S2/PF-03/2003-I)



\begin{thebibliography}{20}

\bibitem{frau} S. Frauendorf, Rev. Mod. Phys. {\bf 73}, 463 (2001).
\bibitem{si03} A. J. Simons et al., Phys. Rev. Lett. {\bf91}, 162501 (2003).
\bibitem{zh01} S. Zhu et al., Phys. Rev. C {\bf64}, 041302(R) (2001).
\bibitem{da05}P. Datta et al., Phys. Rev. C {\bf71}, 041305 (2005).
\bibitem{ch10}D. Choudhury et al., Phys. Rev. C {\bf82}, 061308(R) (2010).
\bibitem{ro11}S. Roy et al., Phys Lett. B {\bf694}, 322 (2011).
\bibitem{ch01} C.J. Chiara, {\it et al.}, Phys. Rev. C {\bf64}, 054314 (2001).
\bibitem{de09} A.Y. Deo {\it et al.}, Phys. Rev. C {\bf79}, 067304 (2009).
\bibitem{pa10}R. Palit {\it et al.}, Nucl. Phys. {\bf A834}, 81c (2010).
\bibitem{rmc} R.M. Clark and A. O.Macchiavelli, Annu. Rev. Nucl. Part. Sci. {\bf 50}, 1 (2000) and references therein.
\bibitem{104Rh} C. Vaman {\it et al.}, Phys. Rev. Lett. {\bf 92}, 032501 (2004).
\bibitem{prm} T. Koike {\it et al.}, Phys. Rev. Lett. {\bf 93}, 172502 (2004).
\bibitem{ko10}T. Koike, Nucl. Phys. {\bf A834}, 36c (2010).
\bibitem{me06} J. Meng, J. Peng, S. Q. Zhang, and S. G. Zhou, Phys. Rev. C {\bf73}, 037303 (2006).
\bibitem{he10}C.Y. He et al., Phys. Rev. C {\bf 83}, 024309 (2010).
\bibitem{he09}C.Y. He et al., Nucl. Phys. {\bf A834}, 84c (2010).
\bibitem{ne10}D. Negi et al., Phys. Rev C {\bf81}, 054322 (2010).
\bibitem{frau93} S. Frauendorf, Nucl. Phys. {\bf A557}, 259 (1993).
\bibitem{mu10} S. Muralithar et al., Nucl. Instrum. Methods Phys. Res. Sect. A {\bf 622}, 281 (2010).
\bibitem{bh01} R. K. Bhowmik, S. Muralithar, and R. P. Singh, DAE Symp. Nucl. Phys. B {\bf44}, 422 (2001).
\bibitem{rad95} D. C. Radford, Nucl. Instrum. Methods A {\bf361}, 297 (1995).
\bibitem{Kr89} A. Kr\"{a}mer-Flecken {\it et al.}, Nucl. Instrum. and Methods A{\bf275} 333 (1989).
\bibitem {St99} K. Starosta {\it et al.}, Nucl. Instrum. Methods A \textbf{423}, 16 (1999).
\bibitem {Pa00} R. Palit {\it et al.}, Pramana \textbf{54}, 347 (2000).
\bibitem {la05}S. Lakshmi {\it et al.}, Nucl. Phys. \textbf{A761}, 1 (2005).
\bibitem {wel91} J. C. Wells, ORNL Physics Division Progress, Reprt No. ORNL-6689, September {\bf 30}, (1991).
\bibitem {nor70} L. C. Northcliffe and R. F. Schilling, At. Data Nucl. Data Tables {\bf 7}, 233 (1970).
\bibitem {Jam75} F. James and M. Roos, Comput. Phys.Commun. \textbf{10}, 343 (1975).
\bibitem{li11}C.B. Li et al., Eur. Phys. J. A {\bf 47} 141 (2011).
\bibitem{25} V.M. Strutinsky, Nucl. Phys. {\bf A95}, 420 (1967).
\bibitem{jen99} {D.G. Jenkins {\it et al.}, Phys. Rev. Lett. {\bf 83}, 500 (1999)}.
\end{thebibliography}
\end{document}